\documentclass{emulateapj}

\def\sun{\ifmmode\odot\else$\odot$\fi}

\shorttitle{T-ReCS observations of LIRGs}
\shortauthors{A. Alonso-Herrero et al.}

\begin{document}

\title{High spatial resolution T-ReCS mid-infrared imaging of Luminous Infrared
Galaxies}


\author{Almudena Alonso-Herrero\altaffilmark{1}, 
Luis Colina\altaffilmark{1},
  Christopher Packham\altaffilmark{2}, Tanio D\'{\i}az-Santos\altaffilmark{1},
George H. Rieke\altaffilmark{3}, James T. Radomski\altaffilmark{4}, and Charles M. Telesco\altaffilmark{3}}
\altaffiltext{1}{Departamento de Astrof\'{\i}sica Molecular 
e Infrarroja, Instituto de Estructura de la Materia, CSIC, E-28006 Madrid,
Spain; aalonso, colina, tanio@damir.iem.csic.es}
\altaffiltext{2}{Department of Astronomy, University of Florida, 211 Bryant
  Science Center, P.O. Box 112055, Gainesville, FL 32611-2055; 
  packham, telesco@astro.ufl.edu}
\altaffiltext{3}{Steward Observatory, University of Arizona, 933 N. Cherry
  Avenue, Tucson, AZ 85721; grieke@as.arizona.edu}
\altaffiltext{4}{Gemini Observatory, c/o AURA, Casilla 603, La Serena,
  Chile; jradomski@gemini.edu }

\begin{abstract}

We present diffraction-limited (${\rm FWHM} \sim  0.3\arcsec$) 
Gemini/T-ReCS mid-infrared (MIR: $N$-band or narrow-band at 
$8.7\,\mu$m) 
imaging of four Luminous Infrared Galaxies (LIRGs) drawn from a representative
local  sample. The MIR emission in the central few kpc is strikingly similar
to   that traced by Pa$\alpha$, and generally  consists of bright nuclear
emission and several  compact circumnuclear and/or extranuclear H\,{\sc ii}
regions.  The central MIR emission  is dominated by these powerful H\,{\sc ii}
regions, consistent with the majority of AGN in this local sample of
LIRGs contributing a minor part of the
MIR emission. The luminous circumnuclear H\,{\sc ii} regions detected in
LIRGs  follow the extrapolation of the 
$8\,\mu$m vs. Pa$\alpha$ relation found for M51 H\,{\sc ii}
knots. The integrated central $3-7\,$kpc of galaxies, however, present elevated 
$8\,\mu$m/Pa$\alpha$ ratios with respect to
individual H\,{\sc ii} regions, similar to the integrated values
for star-forming galaxies. 
Our results show that the diffuse $8\,\mu$m emission, not directly related to
the ionizing stellar population, can be as luminous as that from the resolved
H\,{\sc ii} regions. Therefore, calibrations of the star formation 
rate for distant
galaxies should be based on the integrated $8\,\mu$m emission of nearby
galaxies, not that of the H\,{\sc ii} regions alone.



\end{abstract}


\section[]{Introduction}

There is growing interest in using the mid-infrared (MIR) 
emission of infrared (IR) selected distant galaxies as an
indicator of the massive and dusty 
star formation rate (SFR), analogously to the
widely used SFR vs. IR calibration of Kennicutt (1998). 
The unprecedented sensitivity provided by {\it Spitzer} 
observations reveals the good overall
morphological correspondence between the ionized gas (i.e., 
H$\alpha$, Pa$\alpha$) and the MIR ($8-24\,\mu$m) 
emission of nearby galaxies (Helou et
al. 2004; Hinz et al. 2004; Gordon et al. 2004; Calzetti et
al. 2005, CAL05 hereafter). This suggests that the MIR emission could be used as an
accurate SFR indicator, especially for dusty galaxies. 
For instance, CAL05 and Alonso-Herrero et al. (2006, AAH06 hereafter) 
for resolved  H\,{\sc ii} knots in M51  and for local Luminous
Infrared Galaxies (LIRGs\footnote{$L_{\rm IR[8-1000\mu {\rm m}]} =
10^{11}-10^{12}\,{\rm   L}_\odot$, 
see Sanders \& Mirabel (1996).}) 
respectively, found that
the $24\,\mu$m continuum emission is a good local SFR indicator (see also Wu
et al. 2005).
CAL05 and  P\'erez-Gonz\'alez et al. (2006) 
questioned the use of the IRAC 
$8\,\mu$m continuum 
 emission based on the larger scatter of the $8\,\mu$m 
vs. H$\alpha$ (or Pa$\alpha$)
relation  for resolved H\,{\sc ii} regions in nearby galaxies, 
whereas Wu et al. (2005) found a good correlation 
for local star-forming galaxies. This demonstrates the  
need for further investigation of this issue.

Although {\it Spitzer}  provides highly sensitive imaging
 of LIRGs (Mazzarella et al. 2005), it cannot resolve the sizes
of the MIR emitting regions. 
We present the results of a pilot study intended to understand the 
MIR properties of LIRGs at high spatial resolution (tens-hundreds of parsecs), 
using the Thermal-Region Camera Spectrograph
(T-ReCS; Telesco et al. 1998) 
on the 8.1\,m Gemini-South Telescope.   We observed four  LIRGs 
from the representative sample of 30 local ($z<0.017$) LIRGs 
of AAH06 which was drawn from the  
{\it IRAS} Revised Bright Galaxy Sample (RBGS,
Sanders et al. 2003)  such that the Pa$\alpha$ 
($\lambda_{\rm rest} = 1.876\,\mu$m)
emission line could be observed with NICMOS on the 
{\it Hubble Space Telescope} (HST).
Gemini/T-ReCS and the NICMOS NIC2 camera  
provide comparable spatial resolutions, $\sim 0.30\arcsec$ and 
$\sim 0.15\arcsec$, respectively, making them ideal for this kind of study.
Throughout this paper we use $H_0=75\,$km s$^{-1}$ Mpc$^{-1}$, $\Omega_M=0.3$,
and $\Omega_\Lambda=0.7$.

\section{T-ReCS MIR Imaging Observations}
We obtained imaging observations of 
four LIRGs using T-ReCS in September 2005, and March-April 2006. 
T-ReCS has a plate scale of 
0.089\arcsec pixel$^{-1}$ which results in 
a field of view (FOV) of $\sim 28.5\arcsec \times 21.4\arcsec$.
Three LIRGs (NGC~5135, IC~4518W, and NCG~7130) 
were observed with the broad-band filter $N$ 
(central wavelength $\lambda_{\rm c}=10.36\,\mu$m and width 
 at 50\% cut-on/off $\Delta \lambda=5.27\,\mu$m) and 
 NGC~3256 with the narrow-band Si-2 filter 
($\lambda_{\rm c}=8.74\,\mu$m, and $\Delta \lambda=0.78\,\mu$m).
The on-source integration times were NGC~3256: 304\,s,  NGC~5135
and NGC~7130: 608\,s, and IC~4518W: 1216\,s. 
Packham et al. (2005) described the observation procedures and data
reduction. The 
uncertainties of the photometric calibration were $5-15\%$. The 
standard star FWHMs ($\sim 0.30\arcsec$, Table~1) indicate that 
the T-ReCS observations were effectively diffraction limited, implying 
spatial resolutions of $\sim 50-100\,$pc for our LIRGs. 
For details on the {\it HST}/NICMOS data reduction see
AAH06. The T-ReCS and {\it HST}/NICMOS 
images are presented in Fig.~1.

\begin{table*}
\caption{Photometry.}
\setlength{\tabcolsep}{0.025in}
\scriptsize
\begin{tabular}{lcllcccccccccccc}
\hline
\hline
Galaxy & Class & Dist. & $\log L_{\rm IR}$ & 
$f_\nu(12\,\mu{\rm m})$& \multicolumn{2}{c}{\underline{TReCS FWHM}} & 
\multicolumn{2}{c}{\underline{NICMOS FWHM}} &
\multicolumn{3}{c}{T-ReCS $f_\nu$} & 
\multicolumn{3}{c}{\underline{IRAC $f_\nu(8\,\mu{\rm m}$)}} \\
       &  &       &      & & star & nuclear & nuclear & nuclear & 
nucleus & 5\arcsec & 10\arcsec 
& 5\arcsec & 10\arcsec & $19\arcsec \times 19\arcsec$ \\
& & (Mpc)  & (L$_\odot$) & (mJy) & (\arcsec) & (\arcsec) & (\arcsec) & (pc)
& \multicolumn{3}{c}{(mJy)} & \multicolumn{3}{c}{(mJy)}\\
 \hline
NGC~3256 & HII & 35 & 11.56 & 3570 & 0.30 & 0.45 (N) & 0.40 (N) & 69 (N) & 
$230\pm 11$ (N) & $575\pm 30$ & $845\pm 150$ & 462 & 1399 & 2436 \\
     & HII &   &  &  & & 0.45 (S) & 0.40 (S) & 69 (S) & $40\pm 3$ (S) \\
NGC~5135 & Sy2 & 52  & 11.17 & 630 & 0.31 & 0.35 & 0.16 & $\le 40$ & 
80$^{*}\pm 3$  & $243\pm 20$  & $313\pm 60$   & 214$^*$ & 408 & 490 \\
IC~4518W & Sy2  & 70 & 11.13$^{**}$ & 360$^{**}$ &  0.33 & 0.36 & 0.15 & $\le 51$& 
$117^{*}\pm 6$ &
$165 \pm 11$ &  -- & $154^{*}$ & $167^{*}$ & 192 \\
NGC~7130 & L/Sy & 66 & 11.35 & 580 &  0.31 & 0.45  & 0.60 & 190 &
$130\pm 6$ & $205\pm 10$ & $255\pm40$  & 148$^{*}$ & 232$^{*}$ & 372 \\
\hline

\end{tabular}

Notes.--- 
Column~(1): Galaxy. Column~(2): Activity from optical spectroscopy (see
AAH06 for references). Columns~(3), (4), and (5): Distance, 
IR $8-1000\,\mu$m luminosity, and {\it IRAS}
$12\,\mu$m flux density from Sanders et al. (2003). Columns~(6), and (7): 
T-ReCS FWHM of the standard stars, and nuclei. Columns~(8) and (9): FWHM from
the continuum NICMOS images 
of the nucleus in arcsec and parsec, respectively. Column~(10):
nuclear (narrow-band $8.7\,\mu$m or $N$-band, see \S2) T-ReCS flux density. For the
resolved sources the flux density is 
for a $1.4\arcsec$-diameter aperture. Columns~(11) and (12): T-ReCS flux
densities for 5$\arcsec$- and $10\arcsec$-diameter apertures. The errors only
account for the background estimate  uncertainties. Columns~(13), (14), and (15):
IRAC  $8\,\mu$m flux
densities for 5$\arcsec$- and $10\arcsec$-diameter 
apertures, and the
NICMOS FOV  (no aperture corrections have been applied for extended emission).\\
$^{*}$: Aperture correction for unresolved emission applied.\\
$^{**}$: The {\it IRAS} $12\,\mu$m density flux and IR luminosity are for both
the E and W components of IC~4518.
\end{table*}


For the nuclei of NGC~3256 and NGC~7130 as well as for the 
high surface brightness  regions (assumed to be H\,{\sc ii} regions) 
clearly resolved in the  T-ReCS images, 
we performed aperture photometry. 
The determination of the underlying background 
introduces typical MIR photometry uncertainties 
of $5-20\%$, depending on the brightness of the source. 
For the unresolved sources (nuclei of NGC~5135 and IC~4518W) the
aperture corrections to include all the flux from the point source 
were computed from standard star observations.

To compare our T-ReCS data  with other galaxies 
(see \S3.3),  we converted the T-ReCS $N$-band 
flux densities to IRAC $8\,\mu$m ones, convolving {\it Spitzer}/IRS 
spectra with the appropriate filter transmissions. For 
NGC~5135 and NGC~7130 we found
$f_{\nu}({\rm IRAC} \, 8\,\mu{\rm m})/f_\nu({\rm N-band}) = 0.86$
 and 0.76, respectively.  These might be  average values as 
spatial variations of the MIR spectral properties of the central
regions have been observed in nearby galaxies (e.g., Smith et al. 2004). 
For the Seyfert (Sy) nuclei of NGC~5135 and IC~4518W,
and using a 'generic' IRS Sy2 template (Mrk~3, Weedman et al. 2005),
we obtained $f_{\nu}({\rm IRAC} \, 8\,\mu{\rm m})/f_\nu({\rm N-band}) =
0.56$. This ratio varies by up to 0.1dex in Sy galaxies observed by IRS 
(Buchanan et al. 2006). 
The T-ReCS $8.7\,\mu$m flux densities of NGC~3256 are  
approximately equivalent to the IRAC $8\,\mu$m ones. 
We also performed aperture photometry on the IRAC $8\,\mu$m images (Table~1) 
which have a spatial resolution of $2\arcsec$ (Fazio et al. 2004).

\section{Results and Discussion}
\subsection{MIR Structure on scales of tens-hundreds of parsecs}

The morphological
resemblance (see Fig.~1) between the MIR continuum emitting regions 
and the H\,{\sc ii} regions (traced by the Pa$\alpha$ emission) 
on scales of tens-hundreds of 
pc in LIRGs (see also Soifer et al. 2001) is remarkable.
The near-IR (NIR) 
continuum  emission on the other hand is more extended than the MIR emission 
(see Keto et al. 1997), and generally does
not resemble the morphology of the ionized gas (see AAH06). 
As $A_{\rm 1.6\,\mu{\rm m}} \sim 3.3 \times A_N$ (Rieke \& 
Lebofsky 1985), the lack of correspondence between 
the NIR and MIR (and Pa$\alpha$) emitting
regions is probably not a differential extinction effect. Rather, this 
may reflect age differences since the NIR continuum   
is produced mostly by stellar populations older than the ionizing stellar
populations (see e.g., Alonso-Herrero et al. 2002 and D\'{\i}az-Santos et al. 2006), 
even though highly obscured regions  tend to be associated with the 
youngest regions in LIRGs
 (Soifer et al. 2001; D\'{\i}az-Santos et al. 2006).

The nuclei are the brightest MIR sources in all the
galaxies. The Sy2 nuclei of 
NGC~5135 and IC~4518W\footnote{There are two
  Pa$\alpha$ sources oriented north-south in the central region of
  IC~4518W. The Sy nucleus, the 
brightest Pa$\alpha$ source and the dominant
  continuum source at $\lambda \ge 1.6\,\mu$m, is the south one.} 
appear unresolved in the MIR (see Table~1), and this is 
confirmed by the detection of nuclear point sources
(Table~1, and also AAH06) in the NICMOS continuum images. 
The nucleus of NGC~7130 (classified as a Sy or LINER)  
appears marginally resolved in the T-ReCS images, and the NICMOS
continuum images reveal the presence of at least three sources of similar
flux in the
central $\sim 190\,$pc. This suggests that the putative AGN in NGC~7130 
does not dominate (see \S3.2) the $N$-band nuclear emission. 

\begin{figure*}
\vspace{0.5cm}

\includegraphics[angle=-90,width=16.cm]{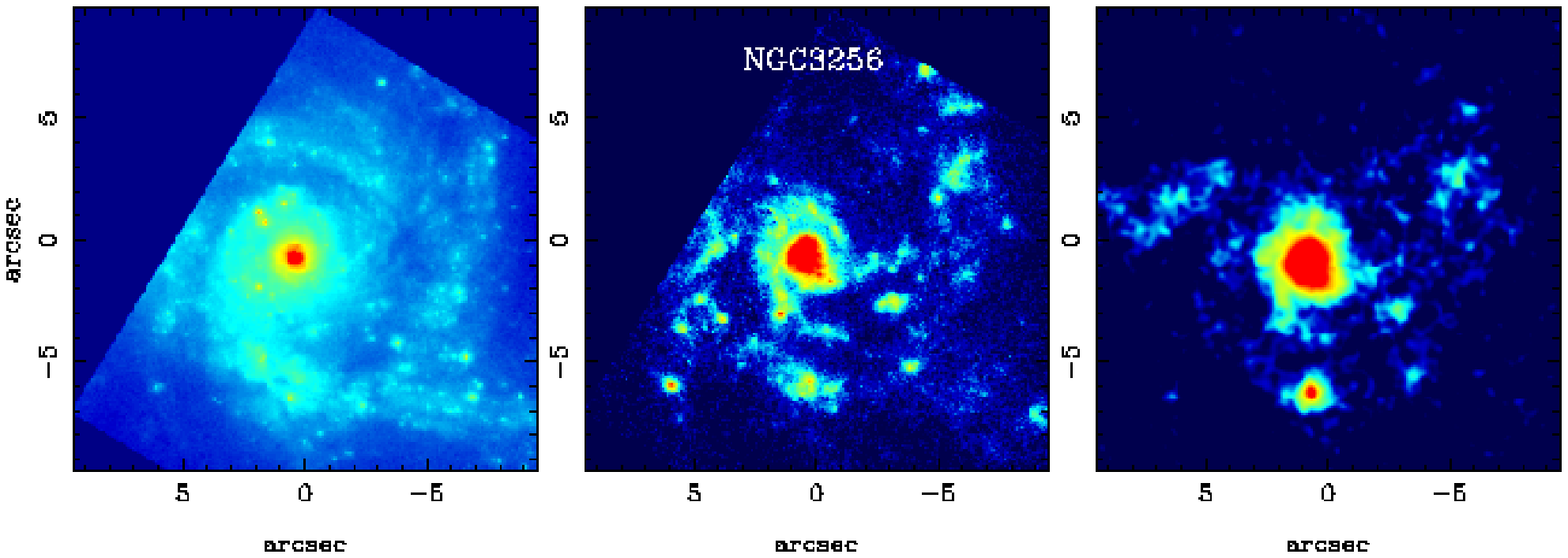}

\vspace{0.5cm}

\includegraphics[angle=-90,width=16.cm]{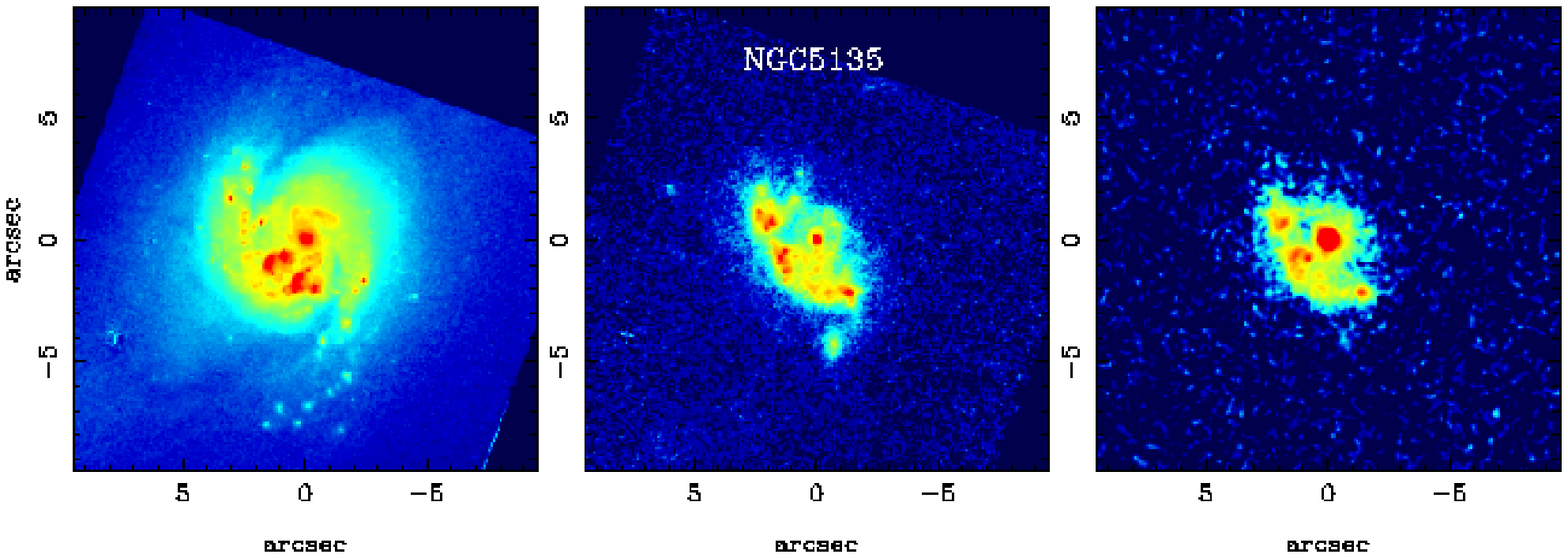}

\vspace{0.5cm}

\includegraphics[angle=-90,width=16.cm]{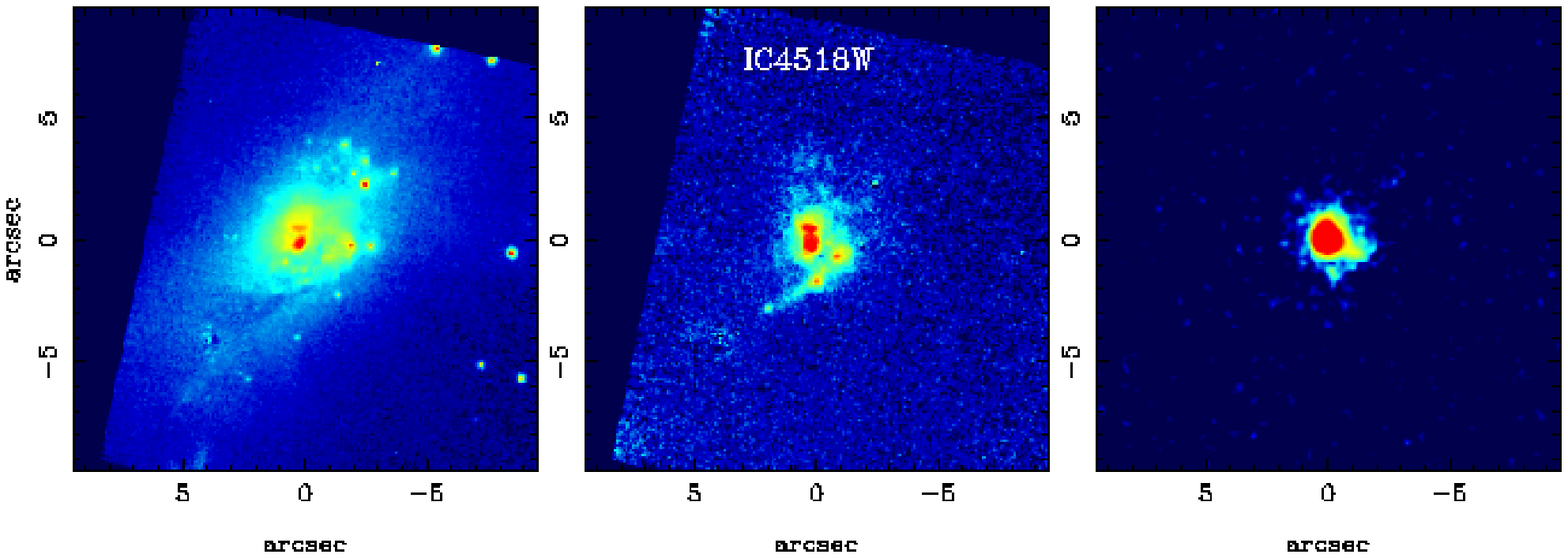}
\vspace{0.5cm}

\includegraphics[angle=-90,width=16.cm]{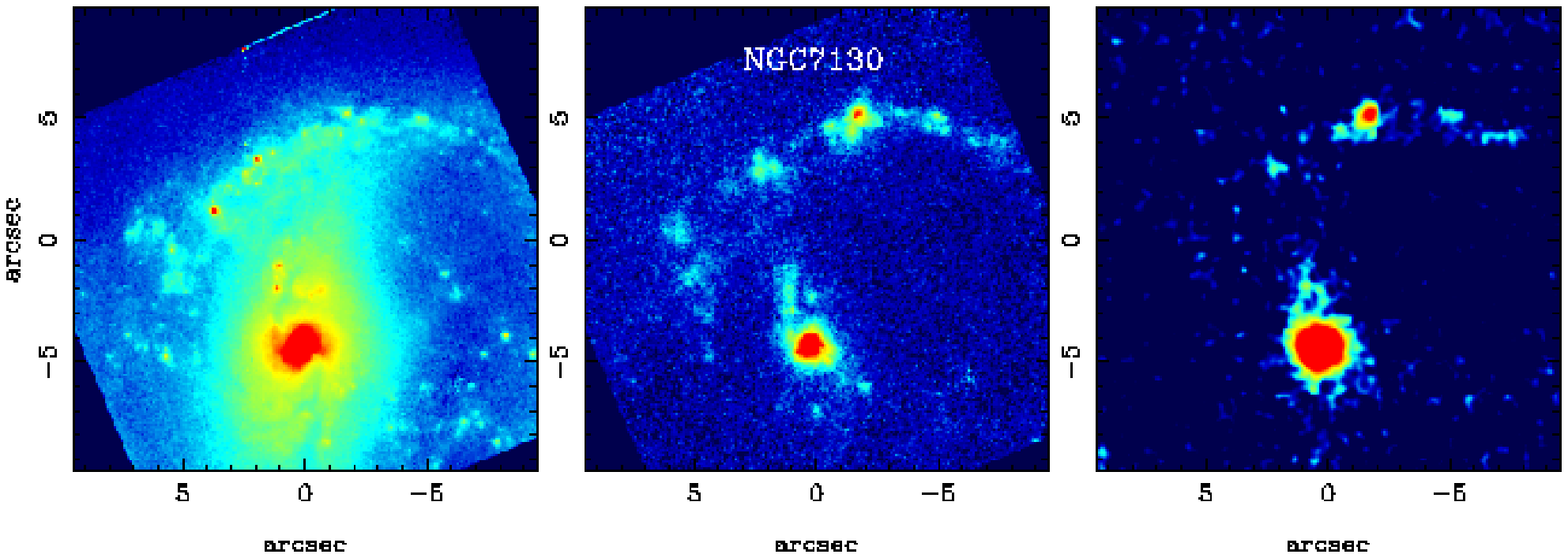}
\caption{{\it Left panels:} {\it HST}/NICMOS $1.1\,\mu$m continuum 
emission images 
except for NGC~3256 which is  at $1.6\,\mu$m 
{\it Middle panels:} {\it HST}/NICMOS continuum-subtracted 
Pa$\alpha$  
line emission. The {\it HST} images except those  of NGC~3256 (see
Alonso-Herrero et al. 2002) are from AAH06. 
{\it Right panels:} Gemini/T-ReCS $N$-band images, 
except for NGC~3256 which is  the narrow-band
$8.7\,\mu$m filter, for approximately the same FOV as the NICMOS imaging. 
The images have been smoothed with Gaussian functions with $\sigma =
1-1.5\,$pixels.
All the NICMOS images
were taken with the NIC2 camera with a 0.076\arcsec pixel$^{-1}$ plate scale. 
Orientation is  north up, and  
east to the left.}
\end{figure*}

\begin{figure}
\vspace{0.5cm}
\includegraphics[angle=-90,width=8.5cm]{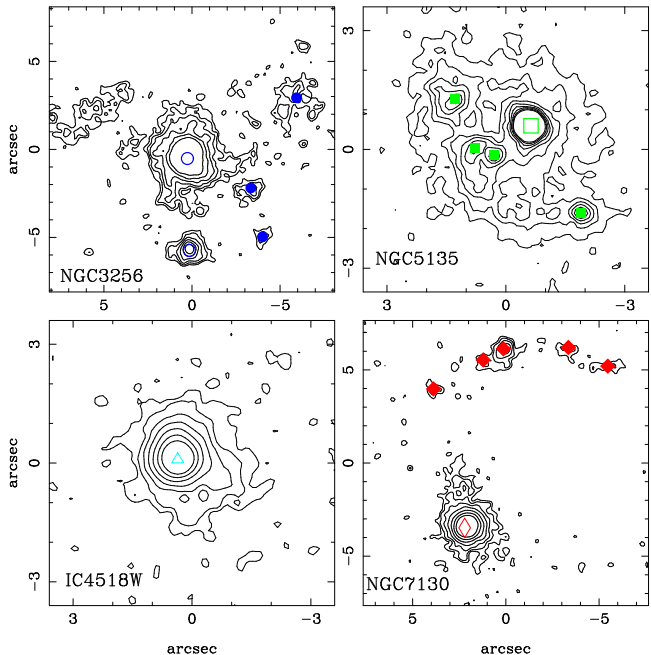}
\caption{The T-ReCS MIR emission contour plots in a logarithmic scale
  except for NGC~5135 for which the scale is linear.
Orientation is north up, east to the left. For each galaxy, the open
  symbols mark the nuclei, whereas the filled symbols indicate the
  positions of high surface brightness regions for which 
  photometry was obtained (see also Fig.~3). For NGC~3256 
we mark the locations of the 
  north and south nuclei, also detected at NIR (Kolitainen et al. 1996) and
  radio (Neff et al. 2003) wavelengths.}
\end{figure}

NGC~3256 is a well-studied merger galaxy, with two bright nuclei (north and
south, see Fig.~2 for location). Kotilainen et al. (1996) and Neff, Ulvestad, \& Campion
(2003) suggested that the nuclei may host low luminosity AGN. 
The NIR (L\'{\i}pari et al. 2000), X-ray  (Lira et al. 2002), and MIR 
spectroscopic
(see Mart\'{\i}n-Hern\'andez et al. 2006) properties  of both nuclei  are 
generally consistent with star formation. 
The extended (${\rm FWHM} \sim 60\,$pc) nature of the NIR and MIR continuum 
emission
of both nuclei further corroborates that such
AGN, if present, are not dominant in the MIR.

\subsection{AGN Contribution to MIR Emission of LIRGs}

Approximately 25\% of local LIRGs host spectroscopically confirmed AGN 
(Veilleux et al. 1995; Goldader et al. 1997; Tran et al. 2001; see Table~1
for our LIRGs).
The Sy2 nucleus of NGC~5135 contributes $\sim 20\,\%$ of the $N$-band
emission within the central 10\arcsec, and even less to the
{\it IRAS} integrated galaxy emission (Table~1).  
The AGN nucleus of NGC~7130, undetected as a point source, is 
responsible for  
$<20\,\%$ of the total MIR emission (see Table~1). 
The MIR emission of IC~4518W is dominated by the 
Sy2 nucleus which accounts for up to $\sim 75\,\%$ (Table~1) 
of the $N$-band flux density of this galaxy 
and approximately half of the MIR emission of the system
(IC~4518W+E). 

Three more galaxies  in the LIRG sample of AAH06, not observed with T-ReCS,
have spectroscopically confirmed Sy nuclei. For two of them, the B1 nucleus of 
the IC~694/NGC~3690 (Arp~299) system
(Garc\'{\i}a-Mar\'{\i}n
et al. 2006), and NGC~7469, the AGN contribution to the MIR emission is 
$\sim 30$\%  (Keto et al. 1997; Genzel et al. 1995; Soifer et al. 2003). 
For the third one, there is no MIR information. Our results are consistent
with the majority of AGN in our local LIRGs not dominating the MIR emission.

\begin{figure}
\includegraphics[angle=-90,width=8.5cm]{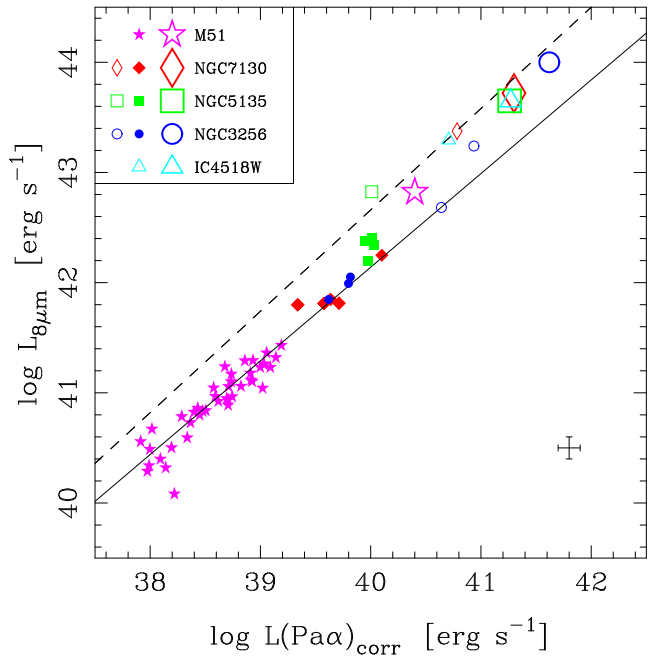}
\caption{Monochromatic ($\nu \ f_{\nu}$) 
$8\,\mu$m vs. extinction-corrected Pa$\alpha$
  luminosities.  
The small open and filled symbols are the nuclei and high
  surface brightness H\,{\sc ii} regions (see Fig.~2) of LIRGs, respectively. 
The photometry for NGC~3256 and NGC~7130 is 
through a $1.4\arcsec$ (240\,pc and 440\,pc respectively) diameter aperture. 
Crowding effects are likely to be present for the photometry of 
the circumnuclear H\,{\sc ii} regions of NGC~5135, even for
the $1.1\arcsec$ \, (270\,pc) aperture used. The error bar
indicates the typical uncertainties associated with 
  the photometry and the 
T-ReCS to IRAC flux density conversion (see \S2). 
The large open symbols for the LIRGs indicate the
  integrated properties over the {\it HST}/NICMOS FOV ($\sim 19\arcsec \times
  19\arcsec$). 
The central 6\,kpc 
M51 H\,{\sc ii} regions (photometry for $\sim 520\,$pc diameter
apertures) and integrated emission from Calzetti et al. (2005) are shown
  as star symbols. The IRAC $8\,\mu$m photometry is 
corrected for extended source emission as described by
  P\'erez-Gonz\'alez et al. (2006).
The solid line is our least-squares fit to the M51 H\,{\sc ii} region data
  from Calzetti et al. (2005) extrapolated to the LIRG luminosities. The
  dashed line is the non-linear fit from Wu et
  al. (2005) for star-forming galaxies
where the H$\alpha$ luminosities have
  been converted to Pa$\alpha$ ones assuming case B recombination.}
\end{figure}

\subsection{The $8-10\,\mu$m MIR Emission as a SFR
  Indicator?} 

Deep surveys at $24\,\mu$m with {\it Spitzer} are detecting a large number
of LIRGs at $z\sim 1$ and above.
These LIRGs contribute significantly to the cosmic 
SFR density and IR background 
at $z\sim 1$ (Le Floc'h et al. 2005; Lagache, Puget, \& Dole 2005; 
P\'erez-Gonz\'alez
et al. 2005). Since the observed $24\,\mu$m
flux densities translate into rest-frame $8\,\mu$m ones at $z = 2$, it is
important to assess the accuracy of the MIR-based SFR indicators.
A number of works (e.g., CAL05; P\'erez-Gonz\'alez et
al. 2006) find that the rest-frame $8\,\mu$m (only dust-emission) 
monochromatic luminosity is not as
tightly correlated with the number of ionizing photons as the $24\,\mu$m
emission (c.f., Wu et al. 2005).

We further explore the
Pa$\alpha$ vs. $8\,\mu$m relation in Fig.~3. 
The LIRG Pa$\alpha$ and MIR emissions are corrected for
extinction using the Rieke \& Lebofsky (1985)
extinction law, and the  $A_V$ averaged over the central emitting regions
(AAH06). For the nuclei of
NGC~3256, we used the nuclear $A_V$.
We assumed that the stellar emission 
contribution at $8\,\mu$m is negligible for our LIRGs.
As can be seen from Fig.~3, although the LIRG  H\,{\sc ii} regions
are up to 10 times more luminous than those in M51, they tend to follow the
extrapolation of the CAL05 relation. 
Conversely, the integrated $\sim 3-7\,$kpc emission of LIRGs deviates 
significantly from the
relation found for individual H\,{\sc ii} regions, but  only  
slightly from the fit of Wu et al. (2005) for star-forming galaxies
(Fig.~3). This could arise 
from two causes. First, Wu et al. (2005) may have underestimated the reddening
(obtained from the Balmer decrements)  in these
dusty star-forming galaxies. Second, the $8\,\mu$m emission may be dominated
by diffuse emission not associated directly with the
H\,{\sc ii} regions (see below).

The LIRG nuclei (except the south nucleus of NGC~3256)  
show elevated MIR/Pa$\alpha$ ratios when compared to H\,{\sc ii} regions. 
Such spatial differences in the MIR emission of 
nuclear and circumnuclear regions of nearby galaxies have also been
observed (e.g., Smith et al. 2004). One possibility is  that an insufficient 
extinction correction (as nuclear 
extinctions in LIRGs tend to be higher than in extranuclear regions)   
 will produce a differential effect, making the nuclei appear more MIR-luminous. 
The AGN present in three of our LIRGs may also play a role, 
as their continua are produced by dust heated to 
higher temperatures than  H\,{\sc ii} regions,  and do not 
present the strong PAH emission characteristic of H\,{\sc ii}+photodissociation regions
(e.g., Laurent et al. 2000; Roche et al. 2006).

The similar behavior of the $8\,\mu$m vs. Pa$\alpha$ relation
for a variety of H\,{\sc ii} regions in M51 and LIRGs
suggests that the $8\,\mu$m emission
is well characterized by a thermal continuum plus PAH features
with no strong variations over the range of conditions
probed here (e.g., metallicity near or over solar, see
AAH06 and CAL05). However, the integrated
central $3 - 7\,$kpc $8\,\mu$m vs. Pa$\alpha$ emission differs
significantly in all these environments from the relation found for
the individual H\,{\sc ii} regions. This may be explained by the presence, in
addition to the bright and compact 
H\,{\sc ii} regions, of a  more diffuse
and extended $8\,\mu$m 
component (see Figs.~1 and 2, and Helou et al. 2004). This extra 
emission at $8\,\mu$m  would be produced not by local,
strong  ionizing sources, but by the diffuse radiation field (see 
e.g., Tacconi-Garman et al. 2005) that permeates
the ISM. As such, the spectrum in these
regions would be characterized by a weak continuum and strong PAH
features with a large equivalent width. Hence when compared to 
individual H\,{\sc ii} regions, an excess of $8\,\mu$m/Pa$\alpha$ emission can
be expected for the integrated properties over a few kpc. 
This is supported by the fact that the
central  $3 - 7\,$kpc emission of LIRGs falls only slightly below
the H$\alpha$  vs. $8\,\mu$m relation found for the integrated
properties of the galaxies studied by Wu et al. (2005), as
indicated in Fig.~3.
For a galaxy with $L_{8\mu{\rm m}}=10^{43}\,$erg s$^{-1}$, the $8\,\mu$m 
vs. Pa$\alpha$ relation for H\,{\sc ii} regions alone predicts
$L({\rm Pa}\alpha$) (and thus SFR) a factor
of $3-4$ larger than that given by the relation found for the integrated
properties of nearby galaxies.

\section{Summary}

The $N$-band (also, the narrow-band at $8.7\,\mu$m) emission  of local LIRGs 
resembles the nuclear and 
H\,{\sc ii} region  emission, as traced by Pa$\alpha$, on scales of
tens-hundreds of parsecs. The AGN contribution to the observed MIR emission 
in our sample of LIRGs is generally small ($<20-30\%$). 
The luminous circumnuclear H\,{\sc ii} regions of LIRGs 
provide evidence that the $8\,\mu$m vs. Pa$\alpha$ relation found for 
M51 knots by CAL05 may extend for a further order of magnitude.
The central
$3-7\,$kpc regions of LIRGs present 
elevated $8\,\mu$m/Pa$\alpha$ ratios with respect to
individual H\,{\sc ii} regions, but similar to those of 
star-forming galaxies (see Wu et al. 2005). This is probably due to the
presence of an extended, 
diffuse $8\,\mu$m component not directly related to the ionizing stellar
population.  A better understanding of the  Pa$\alpha$ (or H$\alpha$) 
vs. $8\,\mu$m relation for
larger samples of nearby star-forming galaxies and LIRGs is 
required before using the IRAC $8\,\mu$m emission as a SFR tracer for
IR-selected high-$z$ 
galaxies.

$\,$
Support was provided by 
the Spanish PNE (ESP2005-01480) and the NSF (0206617).
Based on observations obtained at the Gemini Observatory, which is operated by 
AURA, Inc., under a cooperative agreement
with the NSF on behalf of the Gemini partnership:  NSF (USA),  
PPARC (UK), NRC (Canada), CONICYT (Chile), ARC
(Australia), CNPq (Brazil) and CONICET (Argentina).
Based on observations with the NASA/ESA {\it HST}, 
obtained at the STScI, which is operated by AURA, Inc., under NASA contract
NAS 5-26555.


\begin{thebibliography}{99}


\bibitem{alonso}
Alonso-Herrero, A., et al.  2002, AJ,
124, 166

\bibitem[Alonso-Herrero et al. 2006]{AAH06a}
Alonso-Herrero, A., et al. 2006, ApJ, 650, 835 (AAH06)

\bibitem[Buchanan et al.]{bu06}
Buchanan, C. L. et al. 2006, AJ, 132, 401

\bibitem{CalzettiM51}
Calzetti, D., et al. 2005, ApJ, 633, 871 (CAL05)


\bibitem{diaz}D\'{\i}az-Santos, T., et al. 2006, ApJ, submitted



\bibitem{fazio}Fazio, G. G., et al. 2004, ApJS, 154, 10

\bibitem{Garcia}
Garc\'{\i}a-Mar\'{\i}n, M., et al. 2006, ApJ, 650, 850

\bibitem{Genzel}Genzel, R., et al. 1995, ApJ, 444, 129


\bibitem{b34}Goldader, J. D., et al. 1997, ApJ, 474, 104

\bibitem{Gordon}
Gordon, K. D., et al. 2004, ApJS, 154, 215


\bibitem{Helou}
Helou, G., et al. 2004, ApJS, 154, 253

\bibitem{Hinz}
Hinz, J. L., et al. 2004, ApJS, 154, 259

\bibitem[Kennicutt 1998]{Kenni98}
Kennicutt, R. C. Jr. 1998, ARA\&A, 36, 189

\bibitem{b39}Keto, E., et al. 1997, ApJ, 485, 598



\bibitem{Koti}Kotilainen, J. K., et al. 1996, A\&A, 305, 107

\bibitem{Lagache}
Lagache, G., Puget, J.-L., \& Dole, H. 2005, ARA\&A, 43, 727

\bibitem{Laurent}
Laurent, O., et al. 2000, A\&A, 359, 887

\bibitem{LeFloch}
Le Floc'h, E., et al. 2005, ApJ, 632, 169

\bibitem{lipari}L\'{\i}pari, S., et al. 2000, AJ, 120, 645

\bibitem{Lira}Lira, P., et al. 2002, MNRAS, 330, 259



\bibitem{martin}Mart\'{\i}n-Hern\'andez, N. L., et al. 2006, A\&A, 455, 853

\bibitem{Mazzarella}Mazzarella, J. M., et al. 2005, AAS, 207, 2106

\bibitem{Neff}Neff, S. G., Ulvestad, J. S., \& Campion, S. D. 2003, ApJ, 599,
  1043


\bibitem[Packham et al. 2005]{Packham2005}
Packham, C., et al. 2005, ApJ, 618, L17

\bibitem{perez05}
P\'erez-Gonz\'alez, P. G., et al. 2005, ApJ, 630, 82

\bibitem{perez}
P\'erez-Gonz\'alez, P. G., et al. 2006, ApJ, 648, 987

\bibitem{b70}Rieke, G. H., \& Lebofsky, M. J. 1985, ApJ, 288, 618

\bibitem{roche}Roche, P. F., et al. 2006, MNRAS, 367, 1689 

\bibitem[Sanders \& Mirabel 1996]{Sanders96}
Sanders, D. B., \& Mirabel, I. F. 1996, ARA\&A, 34, 749

\bibitem{b78}Sanders, D. B., et al. 2003, AJ, 126, 1607

\bibitem{Smith}Smith, J. D. T., et al. 2004, ApJS, 154, 199


\bibitem[Soifer et al. 2001]{Soifer01}
Soifer, B. T., et al. 2001, AJ, 122, 1213


\bibitem[Soifer et al. 2003]{Soifer03}
Soifer, B. T., et al. 2003, AJ, 126, 143

\bibitem{Tacconi-Garman}Tacconi-Garman, L. E., et al.
2005, A\&A, 432, 91

\bibitem[Telesco et al. 1998]{Telesco}
Telesco, C. M., et al. 1998, Proc. SPIE, 3354, 534 

\bibitem{b91}Tran, Q. D.,  et al. 2001, ApJ, 552, 527

\bibitem{b94}Veilleux, S., et al. 1995, ApJS, 98, 171

\bibitem{b44}Weedman, D. W.,  et al. 2005, ApJ, 633, 706

\bibitem{Wu}Wu, H., et al. 2005, ApJ, 632, L79

\end{thebibliography}
\end{document}